\documentclass[aps, prl, 
amsmath, showpacs, preprintnumbers, twocolumn, amssymb, amsmath, superscriptaddress]{revtex4-1}
\usepackage{graphicx}
\usepackage{mathptmx, textcomp}
\usepackage[latin1]{inputenc}
\usepackage{tabularx}
\usepackage{units}
\usepackage{color}
\usepackage{bm}

\begin{document}
\title{A levitated nano-accelerometer sensitized by quantum quench}
\author{M.\,Kamba}
\affiliation{Department of Physics, The University of Tokyo; Tokyo, Japan}
\author{S.\,Otabe}
\affiliation{Department of Physics, The University of Tokyo; Tokyo, Japan}
\author{K.\,Funo}
\affiliation{Department of Applied Physics, School of Engineering, The University of Tokyo, Tokyo, Japan}
\author{T.\,Sagawa}
\affiliation{Department of Applied Physics, School of Engineering, The University of Tokyo, Tokyo, Japan}
\affiliation{Quantum-Phase Electronics Center (QPEC), The University of Tokyo, Tokyo, Japan}
\author{K.\,Aikawa}
\affiliation{Department of Physics, The University of Tokyo; Tokyo, Japan}

\date{\today}


\begin{abstract}
We realize a nanoscale accelerometer exploiting the nonequilibrium dynamics of a nanoparticle near the quantum ground state. We explore the dynamics after quenching the trapping potential and find that rapid quenching provides an instance at which the sensitivity is enhanced due to the minimized uncertainty in the position. With rapid quenching, the observed sensitivity is in good agreement with a numerical simulation based on the quantum Langevin equation and approaches to the limit given by the quantum Fisher information. Our results open up a pathway to quantum inertial sensing sensitized by exploiting quench dynamics.
\end{abstract}

\maketitle

%Introduction

\section*{Introduction}

Nonequilibrium dynamics associated with the abrupt modification of a system parameter has been of great interest in a wide variety of quantum systems. While great attention has thus far been paid to the quench dynamics of quantum atomic systems~\cite{d2016quantum,bernien2017probing}, quench dynamics of quantum massive objects has also been an important subject~\cite{aspelmeyer2014cavity,brunelli2015out,raeisi2020quench}. A challenge at nanoscale, both experimentally and theoretically, is to understand and to take control over the impact of quantum mechanics on the nonequilibrium dynamics~\cite{breuer2002theory,wiseman2009quantum,gardiner2004quantum}. 

A single nanoparticle levitated in vacuum is an isolated nanomechanical oscillator that can be brought to the quantum regime~\cite{millen2020optomechanics,gonzalez2021levitodynamics}. The capability to observe and manipulate its motion in the quantum regime makes this system ideal for investigating nonequilibrium quantum thermodynamics on a single-particle level in an isolated environment. To date, nonequilibrium quench dynamics of an optically levitated nanoparticle has been studied in a classical regime~\cite{gieseler2013thermal}, while systematic studies near the ground state still remain unexplored. The nonequilibrium dynamics after an abrupt potential modification was a key ingredient to manipulate the uncertainties in the position and momentum~\cite{rashid2016experimental,muffato2024coherent,rossi2024quantum,duchavn2025nanomechanical}, enabling quantum squeezing~\cite{kamba2025quantum}.  

Levitated objects have found indispensable applications in inertial sensing over decades~\cite{everitt2011gravity,kitching2011atomic}. In particular, nano- and micro-particles levitated in vacuum are a promising candidate for the next generation accelerometer~\cite{ranjit2016zeptonewton,hempston2017force,hebestreit2018sensing,monteiro2020force,rademacher2020quantum}. Conventional accelerometers have relied on detecting the acceleration-induced displacement of mechanical oscillators~\cite{middlemiss2016measurement}. Essentially the same strategy can be employed for realizing acceleration sensing with a levitated particle. The presence of a static acceleration $a$ displaces the minimum of a harmonic potential with the oscillation frequency of $\omega$ by  
 \begin{align}
 \delta = a \omega^{-2}.
\label{eq:Displacement}
 \end{align}

\begin{figure}
\centering
\includegraphics[width=1.0\columnwidth]{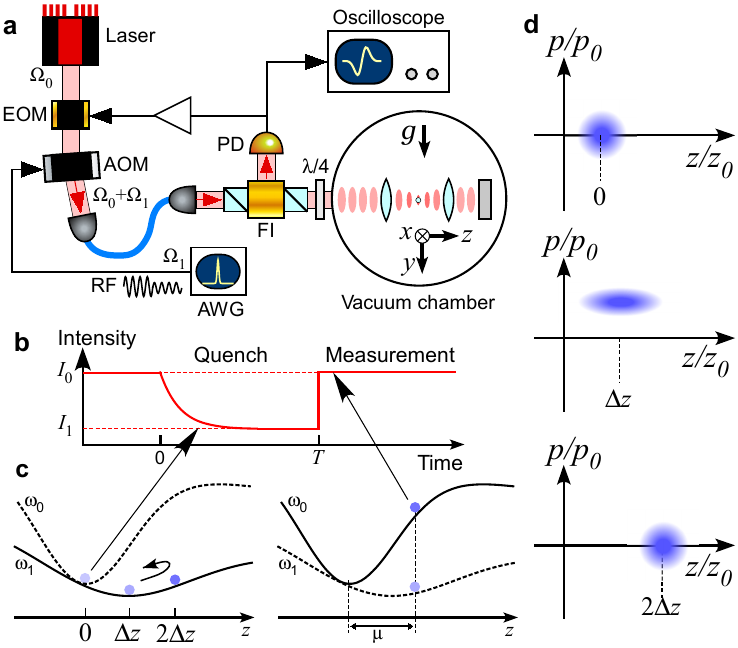}
\caption{{\bf  Overview of the experiments.}
{\bf a,} 
Schematic of our experimental setup. 
 A nanoparticle is trapped in an optical lattice formed by a single-frequency laser.
 The light scattered by the nanoparticle is extracted through a Faraday isolator (FI) and is incident on a photodetector (PD). 
 The signal from the PD is used for both feedback cooling and observing the motion along the optical lattice.
 The intensity of the laser is controlled by an acousto-optic modulator (AOM) driven by radio frequency (RF) from an arbitrary waveform generator (AWG). 
{\bf b,} 
Time sequence of the each cycle of measurements. Feedback cooling of translational motions is turned off just before the quench. 
{\bf c,} 
 Mechanism of an accelerometer with quenching the potential. Due to the static acceleration, the minimum of the harmonic potential displaces with quenching the potential. The displacement induced by the quench $\mu$  is measured by recovering the initial laser intensity at the measurement time $T$.
 {\bf d,} 
 Variation of the uncertainties in the position ($z$) and the momentum ($p$) on the phase space during the quench dynamics. $z_0$ and $p_0$ denote the zero-point fluctuations in the position and the momentum, respectively.
 }
\label{fig:expset}
\end{figure}

In an actual setup, the high oscillation frequency of an optically levitated nanoparticle, which is typically of around 100~kHz and is suitable for ground-state cooling, inevitably makes $\delta$ an extremely small value. Thus, in spite of the remarkably small position uncertainty of the order of 1~pm,  accelerometry with a nanoparticle cooled to near the ground state has been a challenge. Previous demonstrations of gravimetry with a levitated particle are in fact mostly based on micron-sized particles with low oscillation frequencies of less than 100~Hz~\cite{lewandowski2021high,yang2024photon,xu2025compact}, while sensing electric forces with charged nanoparticles has also been successful~\cite{ranjit2016zeptonewton,timberlake2019static,hebestreit2018sensing}. 

In this work, we show that the quench dynamics sensitizes a nanoparticle near the ground state to the projection of the static gravitational acceleration along the direction of its motion.  When $\omega$ is kept constant, the motion of a nanoparticle confined in the potential is insensitive to the acceleration. However, quenching $\omega$ by a factor of about 40 triggers a nonequilibrium dynamics governed by the exerted acceleration (Fig.~\ref{fig:expset}\textbf{b,c}). From the observed dynamics of both the mean position and the position uncertainty, we derive an optimum scheme for quenching and observation. In particular, when the quench is abrupt, we observe substantial oscillations of the position uncertainty, in a similar manner to the behavior crucial for realizing quantum squeezing~\cite{kamba2025quantum}, enabling us to choose an instance at which the measurement fluctuation is minimized and a high sensitivity is achieved (Fig.~\ref{fig:expset}\textbf{d}). A comparison between the observed sensitivity and the theoretically expected limit calculated from the quantum Fisher information (QFI)~\cite{wiseman2009quantum} elucidates that the sensitivity is limited dominantly by background gas collisions. Furthermore, for rapid quenching, the sensitivity calculated with the quantum Langevin equation is in good agreement with experiments. Our work extends previous studies of squeezing the velocity uncertainty and  exploit the nonequilibrium dynamics triggered by an abrupt potential modulation for accelerometry.

\section*{Experimental setup}

Our experimental platform is a charge-neutral silica nanoparticle levitated in an optical lattice formed by a laser with a wavelength of $1551$~nm (Fig.~\ref{fig:expset}\textbf{a}). The nanoparticle has a radius of $\unit[145(2)]{nm}$ and a mass of $m=2.9(2)\times 10^{-17}$~kg.  We investigate the translational motion of the nanoparticle along the optical lattice ($z$ direction) with a frequency of $\omega_0/2\pi=\unit[250(1)]{kHz}$. The position of the nanoparticle is measured by detecting the scattered light with photodetectors and is used for both feedback cooling of its motion and observing the nonequilibrium dynamics. The initial state for the measurement is obtained by applying optical feedback cooling on all the translational motions~\cite{kamba2022optical,kamba2023nanoscale}. The phonon occupation number of the initial state along the optical lattice is $n_z = 0.96(16)$, while the motions in transversal directions are cooled to $n_{x,y} \simeq 14$. The angular motions of the nanoparticle are suppressed by letting it spin around the $z$ axis by introducing a circularly polarized light in the trapping beam~\cite{ahn2018optically,reimann2018ghz,kamba2025quantum}. The pressure of the vacuum chamber for levitating the nanoparticle is kept at around $\unit[3\times 10^{-6}]{Pa}$.

Our apparatus, comprising of an optical setup and a vacuum chamber, is built on an optical table that can be tilted by up to 1~$^\circ$. The optical lattice is nearly perpendicular to the gravity. By tilting the optical table, we can vary the magnitude of the projection of gravity along the optical lattice.  We measure the angle of the table with respect to a horizontal plane with an accuracy of 0.01$^\circ$.

To trigger the dynamics associated with the static acceleration, we switch off cooling and decrease the oscillation frequency by exponentially lowering the intensity of the trapping beam (Fig.~\ref{fig:expset}\textbf{b}). The oscillation frequency is lowered to $\omega_1/ 2\pi = \unit[5.97(1)]{kHz}$ with a $1/e$ time constant of $\tau$. With this quench, the potential minimum displaces by $\Delta z = a(\omega_1^{-2}-\omega_0^{-2})$ (Fig.~\ref{fig:expset}\textbf{c}). In this study, we explore the dynamics for various values of $\tau$ between  $\unit[1]{\mu s}$ and $\unit[50]{\mu s}$.

When the oscillation frequency is decreased to $\omega_1$, the laser intensity is too low to directly reveal the nanoparticle's dynamics with photodetectors. Hence, to measure the position of the nanoparticle in a shallow potential, we abruptly recover the laser intensity to the initial value and observe the oscillation amplitude, from which we extract the absolute value of the displacement from the initial state at each time $T$ (Fig~\ref{fig:expset}\textbf{b}). After the displacement measurements, we reproduce the initial state and repeat the same procedures for about 300 times to derive the distribution of the absolute values of the displacement. The distribution allows us to study the dynamics in terms of both the mean position $\mu $ and its standard deviation $\sigma $ .

\begin{figure}
\centering
\includegraphics[width=1.0\columnwidth]{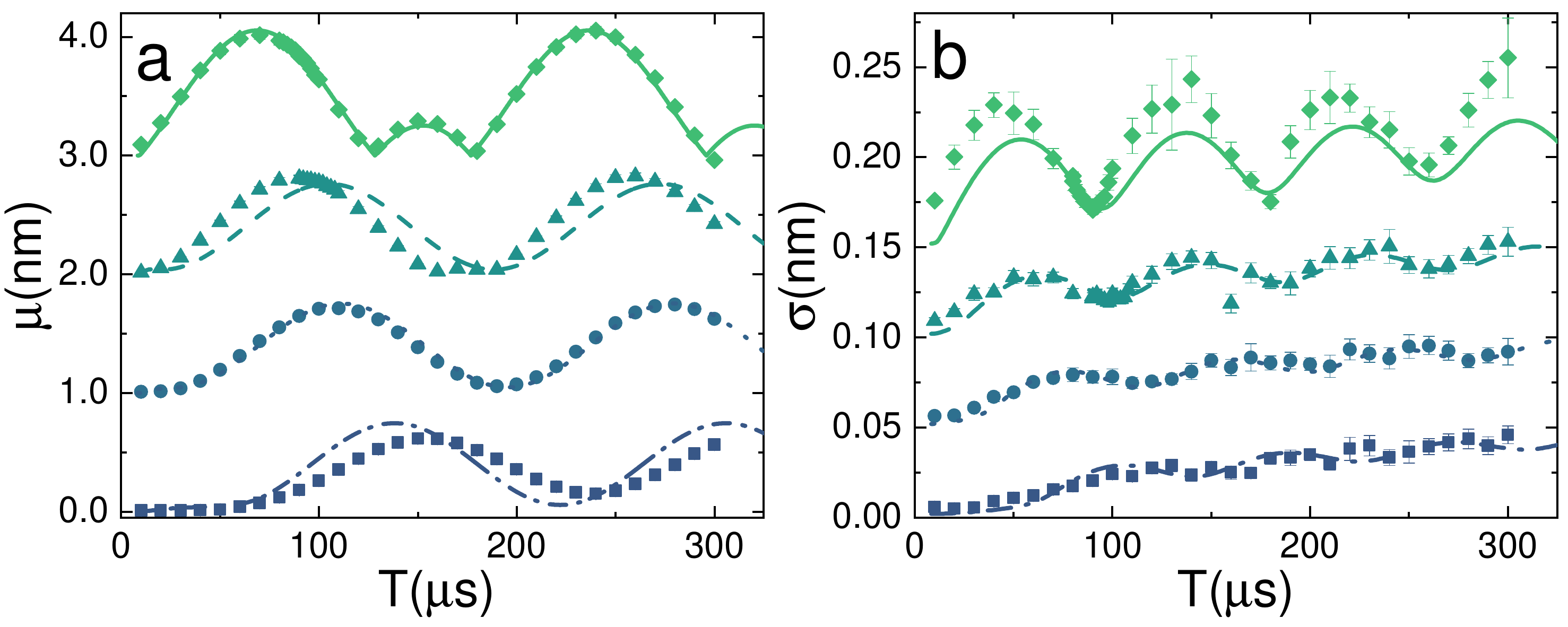}
\caption{ 
{\bf Quench dynamics due to the gravitational acceleration.}  
Each symbol denotes data obtained under the following quench time constants: diamonds, $\tau_1 = \unit[1.95(2)]{\mu s}$; triangles, $\tau_2 = \unit[3.77(3)]{\mu s}$; circle, $\tau_3 = \unit[7.24(6)]{\mu s}$; and square, $\tau_4 = \unit[14.9(1)]{\mu s}$. The solid, dashed, dotted, and dot-dashed lines indicate numerical simulations based on the quantum Langevin equation for $\tau_1, \tau_2, \tau_3$, and $\tau_4$, respectively. 
{\bf a,} 
Oscillatory dynamics of the distribution center $\mu $ after the quench. Each trace is vertically shifted by $\unit[1]{nm}$.
{\bf b,} 
Time evolution of the distribution width $\sigma $ after the quench, exhibiting both coherent breathing-mode oscillations and monotonic increases due to environmental heating. Each trace is vertically shifted by $\unit[0.05]{nm}$.
}
\label{fig:oscillation}
\end{figure}

\section*{Dynamics of the motion after quenching}
 The observed  dynamics of $\mu $ and $\sigma $ is shown in Fig.~\ref{fig:oscillation} for various values of $\tau$. We first investigate the oscillation dynamics of $\mu $. The observed oscillations in $\mu $ is understood as a result of the abrupt shift of the potential minimum (Fig.~\ref{fig:expset}b). In fact, the observed dynamics is in reasonable agreement with our numerical simulation based on the quantum Langevin equation (supplementary information)~\cite{breuer2002theory,gardiner2004quantum,wiseman2009quantum}. By comparing the observed dynamics with numerical simulations, we find that the dynamics of $\mu$ is influenced by the unexpected displacement of the optical potential itself along the light propagation direction, which is triggered by the abrupt decrease in the light intensity. Given that an optical lattice is sensitive to the laser phase, in our numerical simulations, we assumed a phase shift proportional to the intensity with a maximum amplitude of $\unit[37]{pm}$. In this way, the amplitude of the oscillation is well reproduced, while the phase of the oscillation slightly deviates, suggesting that the actual potential shift is more complicated than assumed. Although we have not yet identified the physical origin of such a displacement, an intensity-dependent phase shift and/or thermal effects in some of the optical components, such as an optical fiber, lenses, and mirrors, can cause such an effect. Our results show that the levitated nanoparticle offers a unique opportunity to precisely characterize an intensity-dependent properties of these components.

\begin{figure}
\centering
\includegraphics[width=1.0\columnwidth]{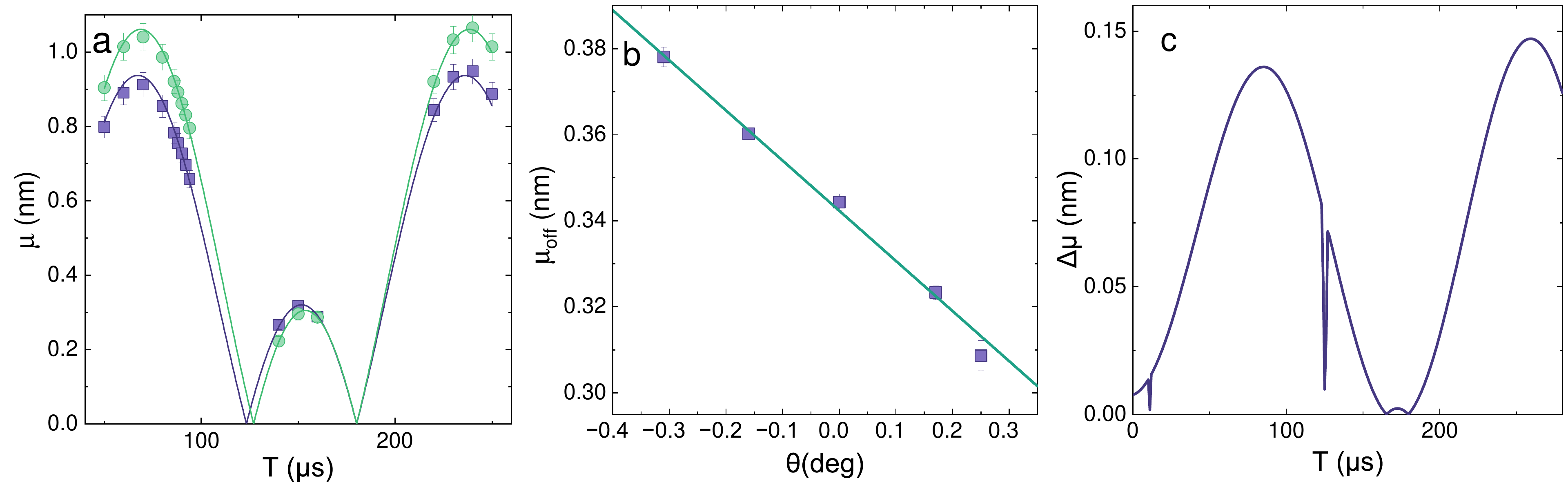}
\caption{
{\bf Dependence of the quench dynamics on the table angle.}  
{\bf a,} 
 Oscillations of $\mu$ for two tilt angles $\theta $ of $\unit[0.25]{^\circ}$ (squares) and $\unit[-0.31]{^\circ}$ (circles) under the shortest quench time of  $\tau_1$.
 The solid lines show fits with Eq.(\ref{eq:oscillation}), from which the offset values shown in {\bf b,}  are extracted.
{\bf b,} 
  The offset values obtained from fits as a function of the tilt angle $\theta $.  
 The measurement is performed with the quench time constant of $\tau_1$.
{\bf c,} 
  The difference between the two curves obtained from fits in {\bf a,}  as a function of $T$. }
\label{fig:DCmeasurement}
\end{figure}

To reveal the impact of the gravitational acceleration exerted on the nanoparticle, we investigate the dynamics for various angles of the optical table $\theta$ for the shortest $\tau$. The projection of gravity along the $z$ direction is given by $a=g\sin (\theta + \theta_{0})$ with $\theta_{0}$ being the intrinsic tilt of the optical lattice with respect to the horizontal plane. As shown in Fig.~\ref{fig:DCmeasurement}\textbf{a}, we clearly observe that the dynamics is dependent on the the angle. By comparing the observed dynamics of $\mu$ with numerical simulations, we determine $\theta_{0}=-3.24^\circ$. 

For a quantitative comparison among various angles, we fit the observed trajectories with a sinusoidal function

\begin{align}
\mu = |A\sin(\omega t + \phi) + \mu_{\mathrm{off}}|,
\label{eq:oscillation}
\end{align}
where $A$ is the amplitude of the oscillation and $\phi$ is the initial phase of the oscillation. We find that the extracted offset values $\mu_{\mathrm{off}}$ clearly exhibit a linear dependence on $\theta$ (Fig.~\ref{fig:DCmeasurement}\textbf{b}), which supports our interpretation that the oscillation is substantially influenced by the displacement of the potential minimum as described in Eq.(\ref{eq:Displacement}).

Our observation shown in Fig.~\ref{fig:DCmeasurement}\textbf{a} implies that the capability of detecting a static acceleration is strongly dependent on the measurement time $T$. To make this point more visible, we show the difference between two trajectories, obtained by fits to the measured dynamics, at two angles, $\unit[0.25(1)]{^\circ}$ and $\unit[-0.31(1)]{^\circ}$, in Fig.~\ref{fig:DCmeasurement}\textbf{c}. We find that the angle dependence oscillates at a frequency of $\omega_1$, suggesting that the position measurement needs to be performed at an optimal time for the sensitive detection of an acceleration. To identify an optimal time sequence, the careful investigation of the uncertainty of the measured position is also crucial. In what follows, we explore the nonequilibrium dynamics of the position uncertainty.

\section*{Time evolution of the position uncertainty}

We now focus on the time evolution of $\sigma $ shown in Fig.~\ref{fig:oscillation}(b). When we take large values of $\tau$ for quenching the potential, we observe a monotonic increase in $\sigma $ suggesting motional heating. Such adiabatic quenching of the potential has been a powerful tool for cooling the ensemble of cold atomic gases~\cite{chen1992adiabatic,kastberg1995adiabatic}. By contrast, when we quench the potential in a short time scale as compared to $2\pi/\omega_1$,  $\sigma $ exhibits an oscillatory dynamics similarly to what has been observed in a study of motional squeezing with a frequency jump~\cite{rashid2016experimental,rossi2024quantum,kamba2025quantum,duchavn2025nanomechanical}.

By comparing the observed values of $\sigma$ with numerical simulations and minimizing the deviations between observed and calculated values of $\sigma$ in the entire time range for all values of $\tau$, we estimate the heating rate to be approximately $\unit[16(2)]{mK/s}$. Given that we fill the chamber with a nitrogen gas during loading the nanoparticle, while the dominant outgassing from the chamber itself is hydrogen, we infer that the residual gas consists mostly of nitrogen and hydrogen.  The observed heating rate is consistent with the gas composition of 60\% of nitrogen and 40\% of hydrogen. Heating rate due to photon recoils is expected to be smaller than these values by more than two orders of magnitude. Thus, we estimate that background gas collisions is a dominant heating mechanism when the potential is quenched.

\begin{figure}
\centering
\includegraphics[width=1.0\columnwidth]{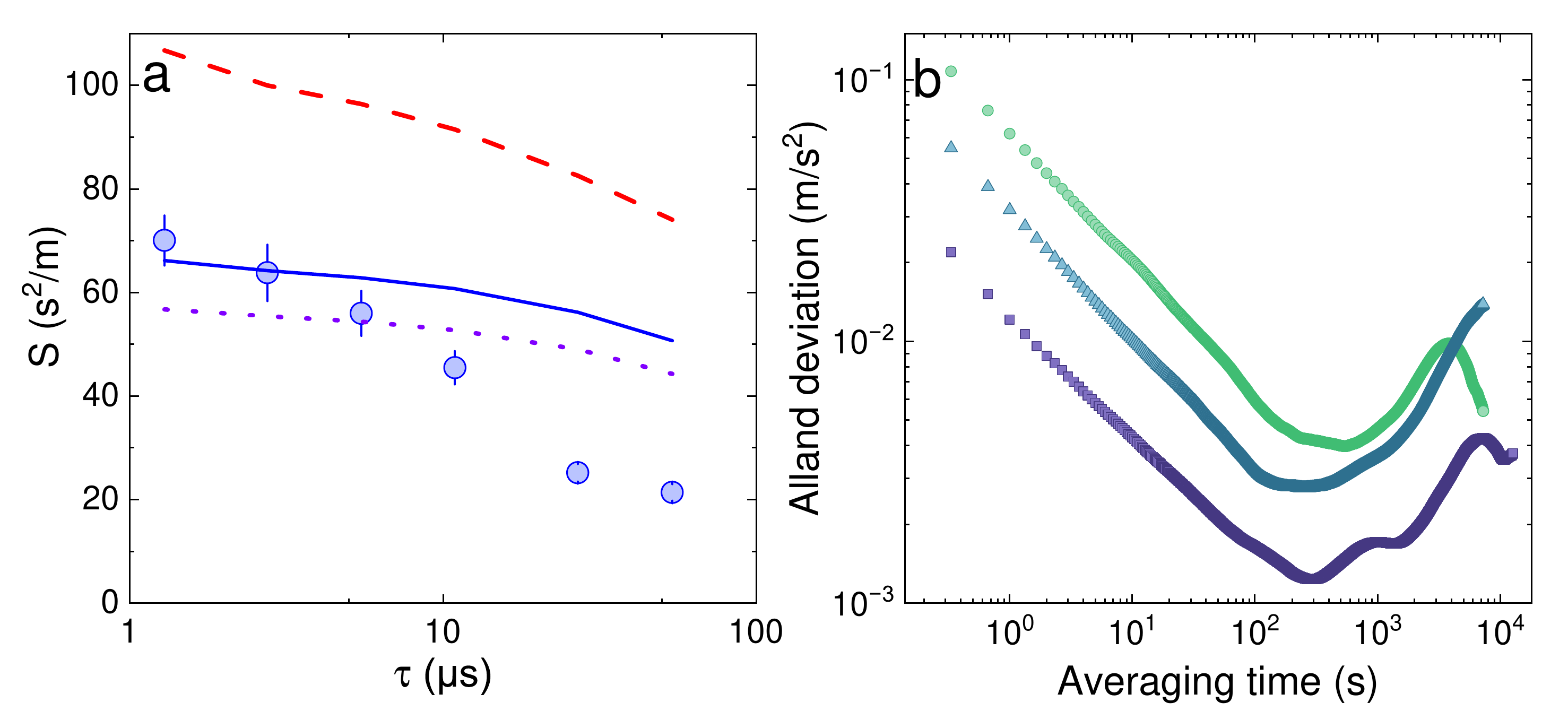}
\caption{
{\bf Variation of the accelerometer sensitivity with the quench time. }   
{\bf a,} 
 The maximum values of $S = (d\mu/da)/\sigma$ as a function of the quench time constant $\tau$. 
 The solid line shows calculated values obtained numerical simulations without fitting parameters with a heating rate of $\unit[16(2)]{mK/s}$,  which is obtained by minimizing the deviation in $\sigma$ between calculated and observed values. For comparison, simulated values for heating rates of $\unit[6]{mK/s}$ and $\unit[22]{mK/s}$, corresponding to two extreme cases that background gases consist of pure hydrogen or pure nitrogen, are also shown by red dashed and purple dotted lines, respectively.  
{\bf b,} 
 Allan deviation of the measured acceleration for three conditions: 
 (1) $\tau_1$ at $T=\unit[89.7]{\mu s}$ (square),
 (2) $\tau_1$ at $T=\unit[46.3]{\mu}$s (circle),
 (3) $\tau_6 = \unit[72.9(5)]{\mu s}$ at $T=\unit[540]{\mu s}$ (triangle). 
 Maximizing the short-term sensitivity is crucial for achieving the high sensitivity for a long time. }
\label{fig:Optimizing}
\end{figure}

\section*{Exploiting nonequilibrium dynamics for achieving optimal sensitivity}

By combining our observation on both $\mu $ and $\sigma $, we explore the limit of the sensitivity as an accelerometer achievable in our scheme. In view of nonequilibrium dynamics, an interesting issue is what time scale of $\tau$ is the best for achieving the highest sensitivity of detecting the acceleration. We define the sensitivity as
\begin{align}
 S = \dfrac{1}{\sigma} \frac{d\mu}{da}, \label{eq:Sensitivity}
 \end{align}
which is a measure of accurately determining the acceleration from a single-shot experiment. Instead of measuring $\mu$ for various angles and for the entire range of $T$, we first identify the optimum measurement time $T_\mathrm{opt}$,
 which gives the largest values of $S$ for each $\tau$ (see appendix for details). We then measure $\mu $ at $T_\mathrm{opt}$ for various angles and obtain the slope $\mathrm{d}\mu /\mathrm{d}\theta $ with a linear fit. Given that $a = g\sin(\theta +\theta_0)\simeq g(\theta +\theta_0)$, we find $\frac{\mathrm{d}\mu }{\mathrm{d}a}\simeq\frac{1}{\,g}\,\frac{\mathrm{d}\mu }{\mathrm{d}\theta }$.

The experimentally obtained values of $S$ as a function of $\tau$ are shown in Fig.~\ref{fig:Optimizing}\textbf{a}. Interestingly, we found that the sensitivity decreases as $\tau$ becomes longer. %From the perspective of the thermodynamic uncertainty relation, one can expect that faster processes would reduce entropy production and thus lead to a better sensitivity because a higher precision requires a larger entropy production~\cite{horowitz2020thermodynamic}. By contrast,
In the context of quantum control, slower (adiabatic) processes result in a higher fidelity and are expected to provide a higher sensitivity, if heating is negligible~\cite{guery2019shortcuts}. Our results show that nonequilibrium dynamics can be exploited to outperform the adiabatic limit, by carefully choosing the measurement time at which the position uncertainty is minimized and the the acceleration-induced displacement takes a large value. 

We further examine the quality of our acceleration measurement by comparing the experimentally obtained sensitivity with the QFI of the system. The QFI provides a fundamental quantum limit on the achievable sensitivity, independent of the specific measurement observable. The sensitivity $S$ defined above corresponds to the square root of the classical Fisher information and is bound by the QFI $F_Q^{(a)}$ as:
   \begin{align}
     S \leq \sqrt{F_Q^{(a)}},
   \end{align}

We calculate the QFI under the assumption of a sudden quench and unitary evolution (see appendix). For the smallest value of $\tau$ with optimal timing, we obtain
   \begin{align}
   \frac{S}{\sqrt{F_Q^{(a)}}} = 9.9(8)\%.
   \label{eq:thsens}
   \end{align}

We estimate that the increased uncertainty due to motional heating after quenching decreases the sensitivity by $9.2(4)$ \%, which is in good agreement with Eq.(\ref{eq:thsens}). The agreement indicates that environmental heating due to background gas collisions is the dominant limitation in the sensitivity and that other decoherence mechanisms such as dephasing and technical noises are negligible. 

Our numerical simulation based on the quantum Langevin equation allows us to directly calculate $S$. The calculated values are also shown in Fig.~\ref{fig:Optimizing}\textbf{a}.  We find that, at quick quenching, the calculated values are in good agreement with the observed values, suggesting that our system is well described by the quantum Langevin equation with heating due to background gas collisions. At slower quenching, the observed values of $S$ deviate from the numerically obtained values. This behavior is due to the smaller values of $\mu$ for larger values of $\tau$ (see supplementary information for details), which imply that a potential displacement, apart from the intensity-dependent phase shift included in our simulations, is present. Exploring the origin of such a displacement will be an important future study for further understanding the optical setup.

   Beyond the fundamental analysis on the quantitative comparison between the experiments and the theory, it is also important to evaluate the practical sensing performance in terms of stability and noise accumulation over a long time. To elucidate the impact of $\tau$ on the sensitivity, we measure the Allan deviation of the acceleration by acquiring the data for about three hours for three conditions with various $\tau$ and $T$. (Fig.~\ref{fig:Optimizing}\textbf{b}).

We find that the high short-term sensitivity is crucial also for the sensitivity on a long term, providing the lowest Allan deviation of approximately $\unit[1.2]{mm/s^2}$. In all conditions, the Allan deviations begin to rise at around five minutes, indicating that the long-term stability of the measurement is limited by the slow drift of the setup. Our results demonstrate that, beyond just maximizing $ S $, the precise control of the measurement time, taking into account the nonequilibrium dynamics, plays a critical role in achieving the long-term sensitivity.

\section*{Conclusions}
  
To summarize, we realize sensing of static gravitational acceleration with a nanoparticle near the ground state by quenching the trapping potential. By observing the dependence on the angle of the optical table, we reveal the nonequilibrium dynamics originates from the projection of the gravity along the measurement direction.  We compare the nonequilibrium quench dynamics for various values of quenching time and find that the sensitivity is maximized at the shortest quenching time. Adiabatic quenching suffers from the increased uncertainties due to environmental heating. The sensitivity achieved with fast quenching is in good agreement with numerical simulations based on the quantum Langevin equation and an estimation based on the QFI. By contrast, the deviation between experiments and calculations at slow quenching reveals that an unexpected potential shift is present in the optical setup.  We anticipate that the sensitivity can be further improved by quenching to an even shallower potential. Decreasing the background pressure is also an important prerequisite to enhance the sensitivity as it is a dominant mechanism of the increased measurement uncertainty. 

The presented scheme of quenching the optical potential has an implication in future quantum mechanical studies where motional heating due to photon scattering needs to be minimized. For revealing the quantum mechanical properties, expanding the wavefunction of a nanoparticle is crucial~\cite{romero2011quantum,stickler2018probing}. Recently, significant efforts have been made to extend the time for free expansion of a levitated nanoparticle with minimum photon scattering~\cite{bonvin2024state,mattana2025trap,steiner2025free,tomassi2025accelerated}. Our results represent an alternative solution to minimize photon scattering simply by decreasing the optical potential. In fact, our work demonstrates that heating after quenching is dominantly due to background gas collisions and photon scattering is not an issue.

Our study shows that the QFI is a useful quantity to characterize the nonequilibrium dynamics of a levitated quantum system in the context of quantum measurements. Realizing the highest sensitivity in our scheme is an intriguing issue beyond sensing applications because such a problem is in close connection with  optimal control relevant to a wide variety of fields~\cite{koch2022quantum,werschnik2007quantum,krotov1996global}. In particular, our setup is ideal for exploring a non-trivial optimization of  optimal control with quantum mechanical behaviors such as quantum squeezing.

%Acknowledgment:\\
\begin{acknowledgments}
 We thank M.\,Kozuma, T.\,Mukaiyama, and M.\,Ueda for fruitful discussions. This work is supported by the Murata Science Foundation, the Mitsubishi Foundation, the Challenging Research Award, the 'Planting Seeds for Research' program, Yoshinori Ohsumi Fund for Fundamental Research, and STAR Grant funded by the Tokyo Tech Fund, Research Foundation for Opto-Science and Technology, JSPS KAKENHI (Grants No. JP16K13857, JP16H06016, JP19H01822, and JP22K18688), JST PRESTO (Grant No. JPMJPR1661), JST ERATO (Grant No.JPMJER2302), JST CREST (Grant No. JPMJCR23I1), and JST COI-NEXT (Grant No. JPMJPF2015). M.\,K. is supported by JSPS (Grant No. JP24KJ1058). K.F. is supported by JSPS KAKENHI (Grants No. JP23K13036 and No. JP24H00831).
\end{acknowledgments}

\section{Supplementary information}

\subsection{Experimental setup}
  A single-frequency laser at a wavelength of $\unit[1551]{nm}$ and with a power of $\unit[215]{mW}$ is focused with an objective lens (NA$=0.85$) and approximately a quarter of the incident power is retro-reflected to form an optical lattice. 
  We load nanoparticles by blowing silica powders placed near the trapping region with a pulsed laser at $\unit[532]{nm}$ at pressures of approximately $\unit[500]{Pa}$. 
  At around $\unit[350]{Pa}$, we apply a positive high voltage to induce a corona discharge and provide a positive charge on the nanoparticle.
  Then we evacuate the chamber with optical feedback cooling for the translational motions.
  We neutralize the nanoparticle via an ultraviolet light at around $\unit[2 \times 10^{-5}]{Pa}$.
  At around $\unit[1 \times 10^{-5}]{Pa}$, we turn off the turbo-molecular pump and the scroll pump, and continue evacuation using an ion pump and a titanium sublimation pump.
 
\subsection{Simulation of particle dynamics}
To understand the time evolution of the nanoparticle after quenching, we numerically simulate its motion along the optical lattice axis. 
The simulation incorporates three main effects: the time-dependent change in the trap frequency, the displacement of the optical potential minimum, and heating due to background gas collisions.
The quantum Langevin equation of motion along the optical lattice is written as~\cite{gardiner2004quantum}: 
\begin{align}
\frac{dz}{dt} =& \frac{p}{m},\\
\frac{dp}{dt} =& -m\omega^2(t) [z-z_{k}(t)] + ma - \gamma p + \xi(t),
\end{align}
where $\omega(t)$ is the time-dependent trap frequency determined by the intensity variation, $a$ is the acceleration, $z_k(t)=\chi(1-\omega^2(t)/\omega_0^2)$ is the displacement of the optical potential minimum, $\gamma$ is the damping rate, and $\xi$ is the thermal noise. We assume that the nanoparticle is heated only by background gas collisions, while we ignore photon recoil heating, which is estimated to be two order of magnitude smaller than heating due to background gas collisions. In the high-temperature limit, the noise correlation function can be approximated by $<\xi(t)\xi(0)>=2m\gamma k_{B} T_0 \delta(t)$, where $T_0$ is the temperature of the background gas~\cite{caldeira1983path,gardiner2004quantum}. 
%Throughout we use $\omega^2(t) \propto I(t)$.
%Note that, relative changes of the table tilt $\theta $ are precisely controlled and measured, the absolute lattice angle $\theta$ is not directly observable.
%Throughout we use $\omega(t)^2 \propto I(t)$.

\subsection{Determination of $T_\mathrm{opt}$}
   To evaluate the sensitivity, we first determined the optimal measurement time $T_\mathrm{opt}$ for each $\tau$.
   For $\tau<\unit[30]{\mu s}$, we used simulations to predict the time dependence of the response $\Delta \mu$ in Fig.~\ref{fig:DCmeasurement}c.
   We found that the time at which the displacement response peaks nearly coincides with the time at which $\sigma $ attains its minimum.
   Thus, $T_\mathrm{opt}$ is determined to be the moment when $\sigma $ is minimized. 
   However, as $\tau$ increases the breathing oscillations in $\sigma $ are gradually suppressed and become a monotonic increase.
   Consequently, the time dependence of the short-term sensitivity is governed primarily by the response slope $\Delta \mu$.
   Hence, for $\tau>\unit[30]{\mu s}$, we use simulations to identify the time that maximizes $\Delta \mu$ and define this time as $T_\mathrm{opt}(\tau)$.

\subsection{Allan deviation analysis}

The measurement time series was acquired with a sampling frequency of
$f_s = \unit[3]{Hz}$. 
For each shot, the post-quench oscillation amplitude $\mathrm{A}_k$ was converted into an effective tilt angle and then into an acceleration value according to
\begin{equation}
    \theta_k = \frac{\mathrm{A}_k}{d\mu/d\theta} + \theta_0, \qquad
    A_k = g\,\theta_k,
\end{equation}
where $d\mu/d\theta$ is the calibration coefficient converting amplitude to angle, $\theta_0$ is the constant misalignment angle between the optical lattice and the table surface, and $g$ is the gravitational acceleration. 

The resulting acceleration series $A_k$ was processed with the MATLAB
\texttt{allanvar} function. 
The estimator was evaluated for averaging factors
$m = 1,2,\dots,\lfloor (N-1)/2 \rfloor$, corresponding to integration times
\begin{equation}
    t_A = \frac{m}{f_s}.
\end{equation}
Thus we obtained the overlapping Allan variance $\sigma_A^2(t_A)$, and the Allan deviation was plotted as
\begin{equation}
    \sigma_A(t_A) = \sqrt{\sigma_A^2(t_A)} ,
\end{equation}
on logarithmic axes. 

\subsection{Heating due to background-gas collisions}
According to the kinetic theory for a particle in a gas, the heating rate of the translational motion is proportional to the background gas pressure $P$~\cite{iwasaki2019electric}:
    \begin{align}
        \Gamma_{BG} = BT_0\frac{P}{R\rho},\ \ B=(4+\frac{\pi}{2})\sqrt{\frac{m_\mathrm{air}}{2\pi k_B T_0}}\label{eq:gamma_Background},
    \end{align}
    where  $\rho$ is the density of the nanoparticle, $m_\mathrm{air}$ is the molecular mass of the background gas, $R$ is the mean radius of the nanoparticle, $k_{B}$ is the Boltzmann constant, and $T_0$ is the temperature of the background gas.
    Using our parameters $R=\unit[145]{nm}$, $\rho=\unit[2260]{kg/m^3}$, $P=\unit[3.0\times10^{-6}]{Pa}$, $T_0=\unit[354]{K}$, and $m_{\mathrm{air}}=\unit[4.65\times10^{-26}]{kg}$, we obtain the heating rate due to background gas $\Gamma_{BG}$ as
    \begin{align}
        \Gamma_{BG} \simeq \unit[22]{mK/s}.
    \end{align}

    In contrast, the measured heating rate is about $\unit[16]{mK/s}$.
    Since $\Gamma_{BG}\propto\sqrt{m_\mathrm{air}}$, a background gas composed entirely of hydrogen ($m_{\rm H_2}\simeq 2\,{\rm amu}$) would yield a heating rate of only $\Gamma_{\rm BG}^{({\rm H_2})}\simeq \Gamma_{\rm BG}^{({\rm N_2})}\sqrt{m_{\rm H_2}/m_{\rm N_2}} \approx \unit[6]{mK/s}$.
    Assuming a mixture of nitrogen and hydrogen, the measured value $\unit[16]{mK/s}$ is reproduced if approximately $\unit[60]{\%}$ of the background gas is nitrogen.
    This fraction is consistent with the fact that outgassing of hydrogen from a stainless steel chamber increases at ultra-high vacuum, and supports the interpretation that the observed heating is due to background gas collisions.

\section{Appendix}

\subsection{Estimation of nanoparticle radius and density}
 The radius and density of the trapped nanoparticle are estimated from the heating rate measurements under different environmental conditions \cite{kamba2023nanoscale}.
 First, we measure the heating rate at around $\unit[5.57]{Pa}$, which is given by the background gas collisions. 
 We measure the pressure with an accuracy of $\unit[0.5]{\%}$ via a capacitance gauge.
 Second, we measure the heating rate at around $\unit[1.2 \times 10^{-6}]{Pa}$, which is determined dominantly by photon recoil heating.
 By combining these results, we determine the radius and the density of the nanoparticle.

\subsection{Stabilization of the angular motion of the nanoparticle}
 To stabilize the angular motions via a gyroscopic effect, we let the nanoparticle spin around the $z$ axis at a frequency of approximately $\unit[8]{GHz}$ by introducing a circularly polarized light into the trapping laser \cite{reimann2018ghz,ahn2018optically}.
 During the measurement, we detune the polarization by $\unit[1.5]{^\circ}$ from the angle for linear polarization to sustain the particle rotation.

\subsection{Position calibration}
 The voltage signal detected from the photodetector was converted into physical displacement of the nanoparticle using a radio-frequency (RF) calibration method~\cite{kamba2025quantum}. 
 In this technique, we apply a controlled frequency modulation to the trapping laser, which shifts the standing-wave potential and thereby induces a displacement of known amplitude at the nanoparticle position. 
 By recording the photodetector response to this forced displacement, we determine the proportionality coefficient between detector voltage and particle displacement.

 \subsection{Derivation of the QFI with quick quench}
We consider the case of frequency jump.
At $t=0$ the state is a thermal state with respect to $\omega_{\mathrm 0}$ (mean phonon number $\bar n$) in the displaced harmonic potential with equilibrium at $z_\mathrm{eq}= a/\omega_{\mathrm 0}^2$.
The initial first and second moments are therefore
\begin{align}
\boldsymbol{\mu}(0)=
\begin{pmatrix}
\langle x(0)\rangle\\[2pt]\langle p(0)\rangle
\end{pmatrix}
=
\begin{pmatrix}
\dfrac{a}{\omega_{\mathrm 0}^2}\\[6pt] 0
\end{pmatrix}, \notag \\
\Sigma(0)=
\begin{pmatrix}
\dfrac{\kappa\,\hbar}{2m\omega_{\mathrm 0}} & 0\\[10pt]
0 & \dfrac{\kappa\,\hbar m\omega_{\mathrm 0}}{2}
\end{pmatrix},
\quad 
\kappa\equiv 2\bar n+1,
\end{align}
where $\Sigma$ is the covariance matrix. In the calculations, we take the value of $\bar n=1.25$ estimated from the time-of-flight measurements.

For $t>0$, the dynamics are governed by the Hamiltonian at $\omega_{\mathrm 1}$ with the same constant force. 
The first moments follow the classical solution of a driven oscillator,
\begin{align}
\langle x(t)\rangle
&= \frac{a}{\omega_{\mathrm 1}^2}
- a\!\left(\frac{1}{\omega_{\mathrm 0}^2}-\frac{1}{\omega_{\mathrm 1}^2}\right)\cos(\omega_{\mathrm 1} t), \\
\langle p(t)\rangle
&= \,m\omega_{\mathrm 1}\,a\!\left(\frac{1}{\omega_{\mathrm 0}^2}-\frac{1}{\omega_{\mathrm 1}^2}\right)\sin(\omega_{\mathrm 1} t).
\end{align}

The covariance matrix is insensitive to the constant-force shift and evolves under the symplectic rotation of frequency $\omega_{\mathrm 1}$:
\[
\Sigma(t)=R_t\,\Sigma(0)\,R_t^{\!\top},\qquad
R_t=
\begin{pmatrix}
\cos\omega_{\mathrm 1} t & \dfrac{\sin\omega_{\mathrm 1} t}{m\omega_{\mathrm 1}}\\[10pt]
-\,m\omega_{\mathrm 1}\sin\omega_{\mathrm 1} t & \cos\omega_{\mathrm 1} t
\end{pmatrix}.
\]
Carrying out the multiplication yields closed-form expressions
\begin{align}
V_{xx}(t)
&=\frac{\kappa\hbar}{2m\omega_{\mathrm 0}}\left[\cos^2(\omega_{\mathrm 1} t)
+\frac{\omega_{\mathrm 0}^2}{\omega_{\mathrm 1}^2}\sin^2(\omega_{\mathrm 1} t)\right],
\\[6pt]
V_{pp}(t)
&=\frac{\kappa\hbar m\omega_{\mathrm 0}}{2}\left[\cos^2(\omega_{\mathrm 1} t)
+\frac{\omega_{\mathrm 1}^2}{\omega_{\mathrm 0}^2}\sin^2(\omega_{\mathrm 1} t)\right],
\\[6pt]
V_{xp}(t)=V_{px}(t)
&=\frac{\kappa\hbar}{4}\!\left(\frac{\omega_{\mathrm 0}}{\omega_{\mathrm 1}}-\frac{\omega_{\mathrm 1}}{\omega_{\mathrm 0}}\right)
\sin\!\bigl(2\omega_{\mathrm 1} t\bigr).
\end{align}

For a single-mode Gaussian state whose dependence on the parameter $a$ enters only through the first moments, the QFI \cite{pinel2013quantum} reduces to
\[
F_Q^{(a)}(t)
=\bigl(\partial_a\boldsymbol{\mu}(t)\bigr)^{\!\top}
\,\Sigma^{-1}(t)\,
\bigl(\partial_a\boldsymbol{\mu}(t)\bigr),
\]
with
\(
\partial_a\boldsymbol{\mu}(t)
=\bigl(f_x(t),\,f_p(t)\bigr)^{\!\top}
\)
given by
\begin{align}
f_x(t)&=\partial_a\langle x(t)\rangle
=\frac{1}{\omega_{\mathrm 1}^2}
-\left(\frac{1}{\omega_{\mathrm 0}^2}-\frac{1}{\omega_{\mathrm 1}^2}\right)\cos(\omega_{\mathrm 1} t),\\
f_p(t)&=\partial_a\langle p(t)\rangle
=\,m\omega_{\mathrm 1}\left(\frac{1}{\omega_{\mathrm 0}^2}-\frac{1}{\omega_{\mathrm 1}^2}\right)\sin(\omega_{\mathrm 1} t).
\end{align}
Using $\det\Sigma(t)=(\kappa\hbar/2)^2$ one may write
\(
\Sigma^{-1}(t)=\dfrac{4}{\kappa^2\hbar^2}
\begin{pmatrix}
V_{pp} & -V_{xp}\\[2pt]
-\,V_{xp} & V_{xx}
\end{pmatrix}.
\)
The QFI takes the compact form
\begin{align}
F_{Q:fast}^{(a)}(t)
=\frac{2m}{\hbar\,\kappa\,\omega_{\mathrm 0}^3}
\Biggl[
1
+\left(\frac{\omega_{\mathrm 0}^2}{\omega_{\mathrm 1}^2}\right)
\!\left(\frac{\omega_{\mathrm 0}^2}{\omega_{\mathrm 1}^2}-1\right)
\bigl(1-\cos\omega_{\mathrm 1} t\bigr)^{\!2}
\Biggr] .
\label{eq:FQ-final}
\end{align}

Based on the experimental parameters used in our optimal sensing protocol, we evaluate the QFI at time $T = T_1/2$ after the quench, where $T_1 = 2\pi/\omega_ \mathrm{1}$ is the oscillation period in the weak trap.
We show that the QFI takes a value of
\begin{align}
F_{Q:fast}^{(a)}(t = T_1/2) \simeq \unit[(4.96 \pm 0.35) \times 10^{5}]{s^4/m^2}.
\end{align}

\subsection{Simulation of particle dynamics}
The quantum Langevin equation of motion along the optical lattice is written as: 
\begin{align}
\frac{dz}{dt} =& \frac{p}{m},\\
\frac{dp}{dt} =& -m\omega^2(t) [z-z_{k}(t)] + ma - \gamma p + \xi(t),
\end{align}
where $\omega(t)$ is the time-dependent trap frequency set by the optical intensity profile, $a$ is the acceleration, $z_{k}(t)$ is the displacement of the optical potential minimum, $\gamma$ is the damping rate, and $\xi$ is the random force.
Throughout we use $\omega^2(t) \propto I(t)$.

\begin{figure}[t]
\centering
\includegraphics[width=0.95\columnwidth]{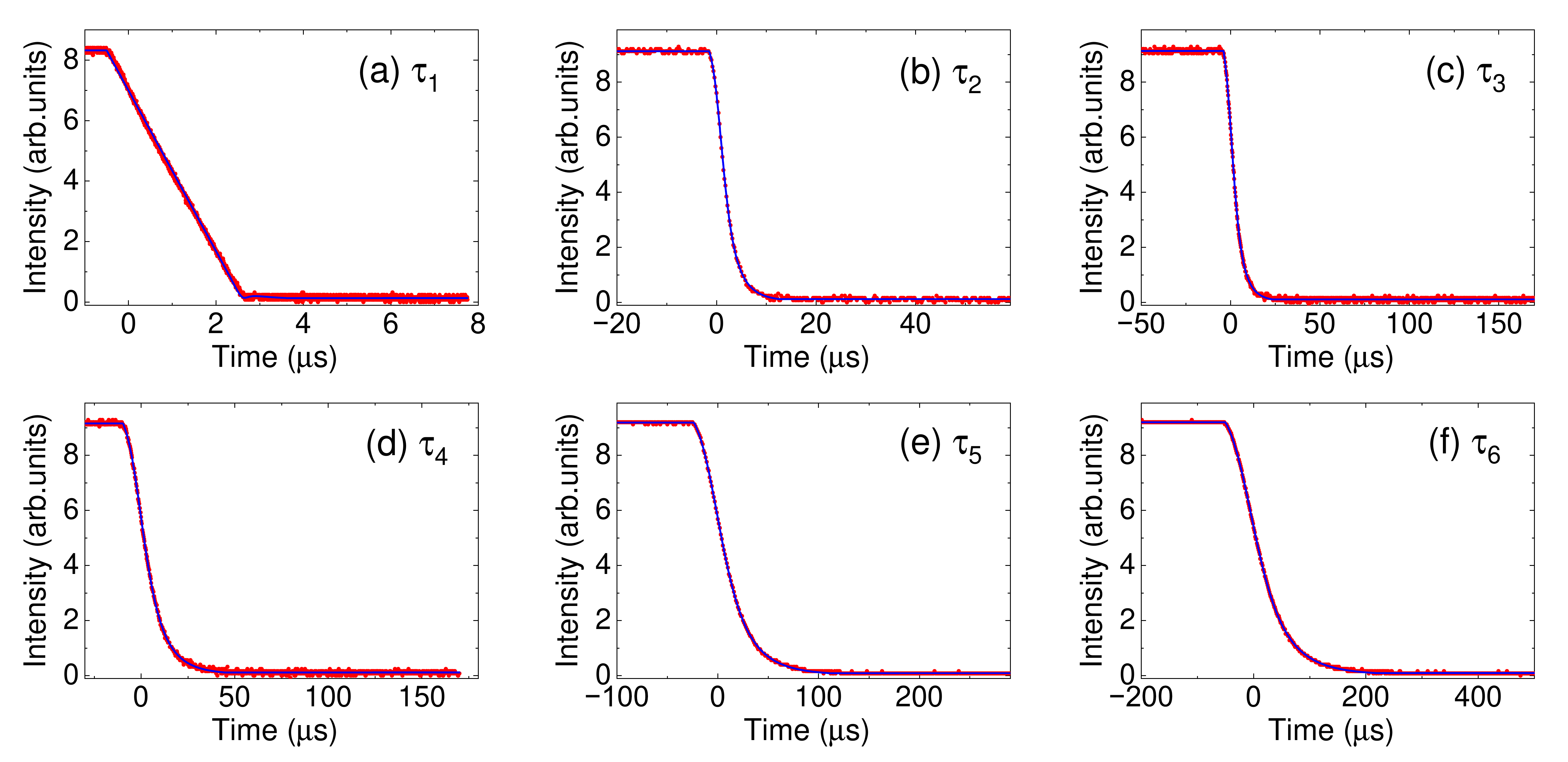}
\caption{
\textbf{Actual intensity variations and fits for them} 
Representative exponential fits used to extract the quench time constant $\tau$. 
Blue dots: measured intensity. Red line: nonlinear least-squares fit to the exponential model. 
Panels (a)-(f) correspond to six independent datasets with different quench conditions. 
The fitted $\tau$ values and their uncertainties are annotated in each panel.
}
\label{fig:tau_fit}
\end{figure}

In the simulations, the intensity of the laser is approximated to be the following form:
\begin{align}
I(t) =& I_\mathrm{lin}(t) \left(1 - R(t)\right) + I_\mathrm{exp}(t) R(t), \\
I_\mathrm{lin}(t) =& I_{0}\frac{q-1}{T_\mathrm{s}}(t-t_0) + I_{0}, \\
I_\mathrm{exp}(t) =& I_{0}\frac{q-1}{1-e^{-5}}\exp\left({-\frac{t-t_0}{\tau_\mathrm{exp}}}\right) + I_0 - I_{0}\frac{q-1}{1-e^{-5}}, \\
R(t) =& \frac{1}{2}\left( 1 + \tanh \left(\frac{t-t_\mathrm{s}}{T_s}\right)\right).
\end{align}
with $t_s, T_s, q, t_0$ being fitting parameters. The fits on the actual intensity variations are shown in Fig.~\ref{fig:tau_fit}.

In order to reproduce the observed dynamics, we introduced an intensity-dependent displacement of the optical potential minimum $z_k(t)=\chi(1-I(t)/I_\mathrm{0})$. 
Here, $\chi$ is the net shift of the optical potential minimum between the initial and final intensities and is determined such that the simulation reproduces the observed dynamics with short quenching. 

From this equation, the following equations are obtained for the mean and covariance.
\begin{align}
\partial_t \langle z \rangle =& \frac{1}{m} \langle p \rangle , \\
\partial_t \langle p \rangle =&  -m\omega^2(t) [\langle z \rangle -z_{k}(t)] + m a - \gamma \langle p \rangle,  \\
\partial_t V_{z} =& \frac{2 C_{zp}}{m}, \\
\partial_t V_{p} =& -2m\omega^2(t) C_{zp} - 2\gamma V_{p} + 2mk_\mathrm{B} T_\mathrm{0} \gamma, \\
\partial_t C_{zp} =& -m\omega^2(t) V_{z} +\frac{1}{m}V_{p} - \gamma C_{zp},
\end{align}
where $T_\mathrm{0}$ is the temperature of background gases.

We simulate the particle dynamics by numerically integrating the above moment equations, with the initial conditions $(\langle z\rangle,\langle p\rangle,V_z,V_p,C_{zp})|_{t=0} = (\frac{a}{\omega_0^2},0,\frac{\hbar}{m \omega_0}(\bar n+\frac{1}{2}),\hbar m \omega_0(\bar n+\frac{1}{2}),0)$. In the calculations, we take the value of $\bar n=1.25$ estimated from the time-of-flight measurements.

\subsection{Histogram model}
In our experiment the readout is the envelope of the oscillation in the high-frequency trap, so each single-shot measurement represents the absolute value of the displacement $|z|$.  
Accordingly, the histogram was fit to the sum of two Gaussians with equal amplitude,
\begin{equation}
\label{eq:fit_model_unnorm}
p_{\rm fit}(z)
= A\,\exp\!\Bigl[-\frac{(z-\mu)^{2}}{2\sigma^{2}}\Bigr]
+ A\,\exp\!\Bigl[-\frac{(z+\mu)^{2}}{2\sigma^{2}}\Bigr],
\qquad z\ge 0,
\end{equation}
where $A$, $\mu$, and $\sigma$ are free parameters.  
We perform nonlinear least-squares fits of the histogram counts to Eq.~\eqref{eq:fit_model_unnorm} to extract $(A,\mu,\sigma)$ for each dataset. 

We model the displacement $z$ as Gaussian.
Therefore, the parameters $\mu$ and $\sigma$ coincide with the mean and standard deviation of the displacement $z$.

\subsection{Validity of the small-angle approximation}
In converting the tilt to acceleration we use $\sin \theta \simeq \theta$ with $\theta=\theta+\theta_0$ (radians).  
Using the Taylor remainder, $\lvert\sin x-x\rvert\le \lvert x\rvert^{3}/6$, the relative truncation error is
\[
\frac{\lvert\sin\theta-\theta\rvert}{\lvert\theta\rvert}\le \frac{\theta^{2}}{6}.
\]
With $\theta_0$ given in the main text and $\theta_{\rm meas}\in[-0.5^\circ,0.5^\circ]$, we have
\(
\lvert\theta\rvert\in[\lvert\theta_0\rvert-0.5^\circ,\;\lvert\theta_0\rvert+0.5^\circ].
\)
For the fitted value $\theta_0\simeq-3.24^\circ$, this yields
\[
\frac{\lvert\sin\theta-\theta\rvert}{\lvert\theta\rvert}\;\le\;\frac{\theta^{2}}{6}
\;<\;7.1\times10^{-4} ,
\]
attained at $\lvert\theta\rvert\approx 3.74^\circ$.  

\subsection{Laser phase noise-induced heating}
We consider a harmonically trapped nanoparticle of mass $m$ oscillating at angular frequency
$\omega_1$ along the trap axis. 
Laser phase noise (LPN) causes the optical lattice nodes to move, producing an effective trap position displacement noise $P_n(t)$.
Let $\tilde P_n^{\,2}(\omega)$ denote the power spectral density (PSD) of that displacement with units $\mathrm{m^2/Hz}$.

For a narrow mechanical resonance, the on-resonance LPN heating rate is given by~\cite{kamba2021recoil} 
\begin{equation}
\label{eq:onres}
\dot E_{\mathrm{LPN}} \;\equiv\; \Gamma_E \;\simeq\; \frac{1}{2}\,m\,\omega_1^{4}\,\tilde P_n^{\,2}(\omega_1).
\end{equation}

The trap-center displacement PSD is obtained from the laser frequency-noise PSD $S_\nu(f)$ (units $\mathrm{Hz^2/Hz}$) evaluated at $f=f_1=\omega_1/2\pi$:
\begin{equation}
\label{eq:freq2disp}
\;
\tilde P_n^{\,2}(2\pi f) \;=\; \Bigl(\frac{\lambda\, d}{c}\Bigr)^{\!2}\, S_\nu(f),\,
\end{equation}
where $\lambda$ is the laser wavelength, $d$ the distance between the nanoparticle and the
retro-reflecting mirror, and $c$ the speed of light.

The displacement-noise PSD at $f_1$ is
\begin{equation}
\tilde P_n^{\,2}(\omega_1)
= \Bigl(\tfrac{\lambda d}{c}\Bigr)^{\!2} S_\nu(f_1)
= 7.37(27)\times 10^{-33}\ \mathrm{m^2/Hz}.
\end{equation}
With $\omega_1=\unit[2\pi\times 5.97(1)]{kHz}$, corresponding values are written as:
\begin{align}
\Gamma_E &=
\frac{1}{2}m\omega_1^{4}\tilde P_n^{\,2}(\omega_1)
= \,2.11(1)\times 10^{-31}\ \mathrm{J/s},\\
\dot{\bar n} &= \Gamma_E/(\hbar\omega_1)
= \,5.3(2)\times 10^{-2}\ \mathrm{s^{-1}},\\
\frac{dT}{dt} &= \Gamma_E/k_B
= \,1.5(1)\times 10^{-8}\ \mathrm{K/s}.
\end{align}

Here, we use following values to calculate: mirror distance $d = \unit[16.6(3)]{mm}$, wavelength: $\lambda = \unit[1551]{nm}$, nanoparticle mean radius: $r = \unit[145(2)]{nm}$, density: $\rho = \unit[2.26(13)\times 10^3]{kg/m^3}$, laser frequency-noise PSD at $f_1=\omega_1/(2\pi)$: $S_\nu(f_1) \simeq \unit[1]{Hz^{2}/Hz}$.

\subsection{Influence of vibration on heating}

In the main text, we reported a heating rate of $\unit[16(2)]{mK/s}$, which is about 300 times larger than the expected photon recoil heating. 
To minimize vibrational noise, all measurements presented in the main text were performed with the turbo-molecular and scroll pumps turned off, and the chamber was maintained only with the ion pump and titanium sublimation pump.

To investigate the possible contribution of mechanical vibrations, we intentionally repeated the measurement while operating the turbo-molecular and scroll pumps. 
Under these conditions, we observed a further increase in the heating rate compared to the pump-off configuration by a factor of about 6 (Fig.~\ref{fig:vibration}). This observation strongly suggests that vibrational noise can couple to the particle motion and act as a dominant source of excess heating.

\subsection{Uncertainty analysis of the data}
We analyze the following source of uncertainties. 

\subsection{1. Uncertainty in the frequency $\omega_0$}
 Without feedback cooling, we record the PSD peak frequency ten times, obtaining $\{f_{\mathrm 0,i}\}_{i=1}^{10}$.  
 We use the sample mean of the ten peak frequencies as the point estimate of $\omega_{\mathrm 0}$ ($\omega_{\mathrm 0}=2\pi\times \bar f_{\mathrm 0}$) and the sample standard deviation as its uncertainty ($\delta\omega_{\mathrm 0}=2\pi\times s_{f_{\mathrm 0}}$).

\subsection{2. Uncertainty in the frequency $\omega_1$}
 At the shortest intensity-quench time constant, we measure the $\mu$ oscillation under seven table-tilt settings and fit each trace to a sinusoid to obtain
 $\{f_{\mathrm 1,j},\,\delta \hat f_{\mathrm 1,j}\}_{j=1}^{7}$. 
 We use the sample mean of the seven frequencies from the fitting of $\omega_{\mathrm 1}$ ($\omega_{\mathrm 1}=2\pi\times \bar f_{\mathrm 1}$) and combine the fit uncertainties by adding them in quadrature and the standard deviation as its uncertainty ($\delta\omega_{\mathrm 1}=2\pi\times s_{f_{\mathrm 1}}$).

\subsection{3. Uncertainty of the QFI}
 Using Eq.~\eqref{eq:FQ-final} at $t=T_1/2=\pi/\omega_{\mathrm 1}$, the QFI reduces to
 \begin{align}
 F_Q^{(a)}\!\left(t=\tfrac{T_1}{2}\right) =\frac{2m}{\hbar\,\kappa\,\omega_{\mathrm 0}^3}\,\Bigl[1+4\,r(r-1)\Bigr],
 \quad \notag \\
 \kappa\equiv 2\bar n+1,\quad r\equiv\frac{\omega_{\mathrm 0}^2}{\omega_{\mathrm 1}^2}.
 \end{align}

 Assuming small, uncorrelated uncertainties in $p=\{m,\bar n,\omega_{\mathrm 0},\omega_{\mathrm 1}\}$,
 the relative uncertainty is
 \begin{align}
 \left(\frac{\delta F_Q}{F_Q}\right)^{\!2}\simeq
 \left(\frac{\delta m}{m}\right)^{\!2}
 +\left(\frac{\delta\kappa}{\kappa}\right)^{\!2} \notag \\
 +\left[-3+\frac{8\,r(2r-1)}{\,1+4r(r-1)\,}\right]^{\!2}
 \left(\frac{\delta\omega_{\mathrm 0}}{\omega_{\mathrm 0}}\right)^{\!2} \notag \\
 +\left[\frac{8\,r(2r-1)}{\,1+4r(r-1)\,}\right]^{\!2}
 \left(\frac{\delta\omega_{\mathrm 1}}{\omega_{\mathrm 1}}\right)^{\!2},
 \end{align}
 with $\frac{\delta\kappa}{\kappa}=\frac{2\,\delta\bar n}{\,2\bar n+1\,}$.

\subsection{4. Uncertainty of the mean displacement $\mu $}
 The mean displacement $\mu $ is obtained from folded-normal distribution fits to the position histograms. 
 The statistical uncertainty $\delta\mu$ is given by the standard error of the fit. 
 In addition, the voltage-to-displacement calibration factor introduces a systematic uncertainty. 
 The total error bars for $\mu $ shown in Figs.~\ref{fig:oscillation}\textbf{a}, \ref{fig:DCmeasurement}\textbf{a}, and \ref{fig:DCmeasurement}\textbf{c} are obtained by adding the statistical and calibration contributions in quadrature.

\subsection{5. Uncertainty of the distribution width $\sigma $}
 The distribution width $\sigma $ is also obtained from folded-normal distribution fits. 
 The statistical uncertainty $\delta\sigma $ is given by the standard error of the fit. 
 In addition, the voltage-to-displacement calibration factor introduces a systematic uncertainty. 
 The total error bar for $\sigma $ shown in Figs.~\ref{fig:oscillation}\textbf{b} is obtained by adding the statistical and calibration contributions in quadrature.

\subsection{6. Uncertainty of the measured tilt angle $\theta $}
 The tilt angle $\theta$ of the optical table is measured with a digital protractor. 
 The resolution limit of the device introduces an angular uncertainty of $\pm \unit[0.01]{^\circ}$, which determines the horizontal error bars in Fig.~\ref{fig:DCmeasurement}\textbf{b}.

\subsection{7. Uncertainty of the fitted offset $\mu_{\mathrm{off}}$ in Fig.~\ref{fig:DCmeasurement}\textbf{b}}
 For each tilt angle we extract the offset parameter $\mu_{\mathrm{off}}$ by fitting the 
 oscillation traces with the rectified sinusoidal model $|\sin(\omega t+\phi)+\mu_{\mathrm{off}}|$.
 The vertical error bars in Fig.~\ref{fig:DCmeasurement}\textbf{b} correspond to the standard errors of this offset parameter obtained  from the fits.

\subsection{8. Uncertainty from the absolute tilt reference $\theta_0$}
 In our calibration of acceleration along the lattice axis, we write
 \begin{align}
   a(\theta) = g\,\sin\!\bigl(\theta  + \theta_0\bigr),
 \end{align}
 where $\theta$ is the tilt of the optical table and $\theta_0$ is a constant offset that accounts for the misalignment between the table reference and the optical lattice axis. 
 We determine $\theta_0$ from a fit to the shortest-$\tau$ data set with the simulation.

\subsection{9. Uncertainty in the angular $d\mu/da$}
 The angular response $d\mu/d\theta $ is extracted from a linear fit of the center $\mu$ vs.\ tilt angle $\theta $.
 The slope uncertainty provides $\delta(d\mu/d\theta )$.
 Here $\theta $ denotes the measured values of the tilt of the optical table, not the lattice tilt. 
 From the simulation described in the main text we determine the constant misalignment angle $\theta_0$ between $\theta $ and the tilt of the optical lattice $\theta$. 
 The effective acceleration projected along the lattice axis $a$ is then
 \begin{align}
   a = g\sin(\theta +\theta_0)\simeq g(\theta +\theta_0).
 \end{align}
 Consequently, the slope with respect to table tilt converts to a slope with respect to acceleration as
 \begin{align}
   \frac{d\mu}{da} = \frac{1}{g\cos(\theta +\theta_0)}\frac{d\mu}{d\theta }\simeq \frac{1}{g}\frac{d\mu}{d\theta }.
 \end{align}

 The corresponding uncertainty of $d\mu/da$ due to  the statistical slope error $\delta\!\left(d\mu/d\theta \right)$ is
 \begin{align}
   \delta\!\left(\frac{d\mu}{da}\right)
   \simeq 
     \frac{1}{g}\,\delta\!\left(\frac{d\mu}{d\theta }\right).
 \end{align}

\subsection{10. Uncertainty of the sensitivity $S(t)$}
 The single-shot sensitivity is defined as
 \begin{align}
   S(t)=\frac{d\mu/da}{\sigma}.
 \end{align}
 Assuming that the uncertainties of $d\mu/da$ and $\sigma$ are uncorrelated, the standard error propagation gives
 \begin{align}
   \delta S
   \simeq
   S
   \sqrt{
     \left(\frac{\delta\!\left(d\mu/da\right)}{d\mu/da}\right)^2
     +
     \left(\frac{\delta\sigma}{\sigma}\right)^2 }.
   \label{eq:S_unc_rel}
 \end{align}

 The width uncertainty $\delta\sigma$ is obtained from the folded-normal distribution fit to the histograms at the analysis time.
 For Fig.~\ref{fig:Optimizing}\textbf{a}, at each $\tau$ we record multiple sets of $N=300$ realizations for several tilt angles. 
 From each histogram $i$ we obtain the width $\sigma_i$ and its standard error $\delta\sigma_i$ from the folded-normal distribution fit. 
 The weighted mean width and its uncertainty are
 \begin{align}
   \bar{\sigma}=\frac{\sum_i \sigma_i/\delta\sigma_i^2}{\sum_i 1/\delta\sigma_i^2},\qquad
   \delta\bar{\sigma}=\sqrt{\frac{1}{\sum_i 1/\delta\sigma_i^2}}.
 \end{align}

\begin{figure}
\centering
\includegraphics[width=0.95\columnwidth]{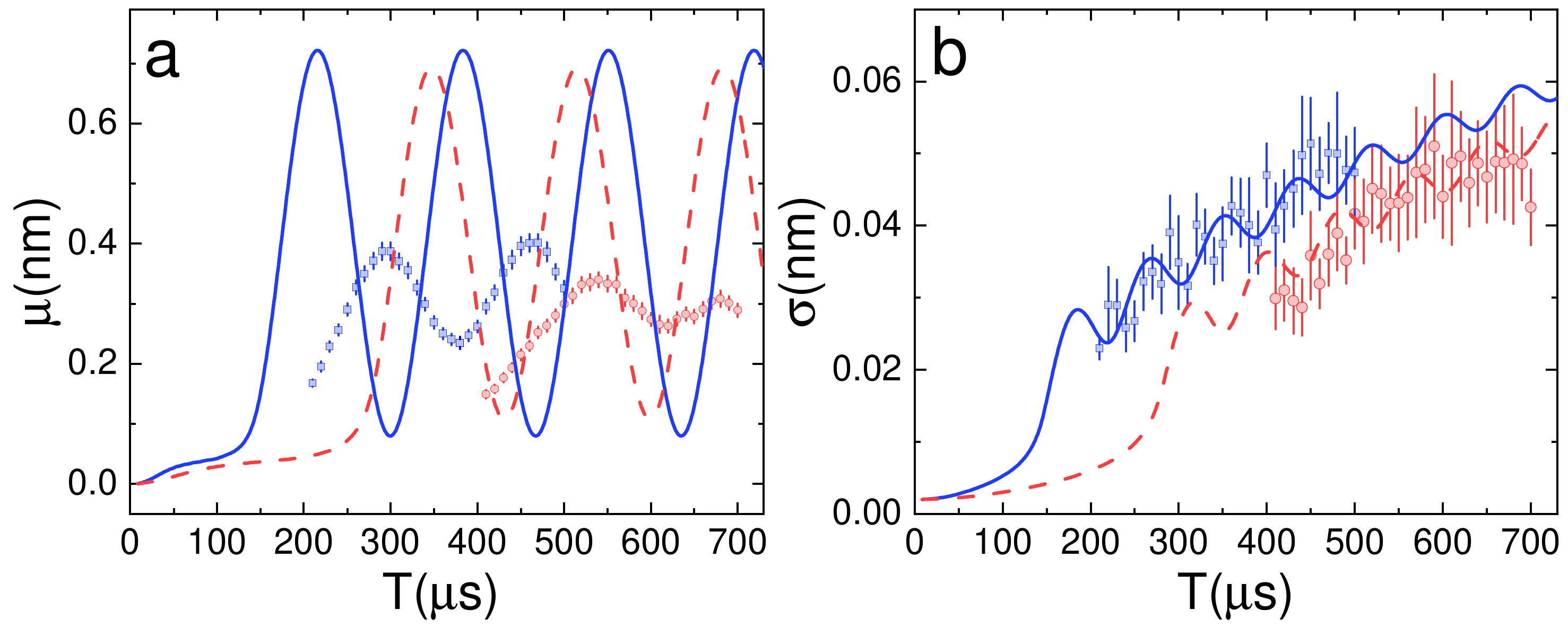}
\caption{
\textbf{Time evolution of $\mu$ and $\sigma$ for slow quenching}
Each symbol denotes data obtained under the following quench time constants: squares, $\tau_5 = \unit[36.6(2)]{\mu s}$; circles, $\tau_6 = \unit[72.9(3)]{\mu s}$; The solid and dashed indicate numerical simulations for $\tau_5$ and $\tau_6$, respectively. The deviation in $\mu$ between experiments and calculations implies a potential drift that is not included in our simulation.
{\bf a,} 
Oscillatory dynamics of the distribution center $\mu $ after the quench. 
{\bf b,} 
Time evolution of the distribution width $\sigma $ after the quench.
}
\label{fig:vibration}
\end{figure}

\begin{figure}
\centering
\includegraphics[width=0.8\columnwidth]{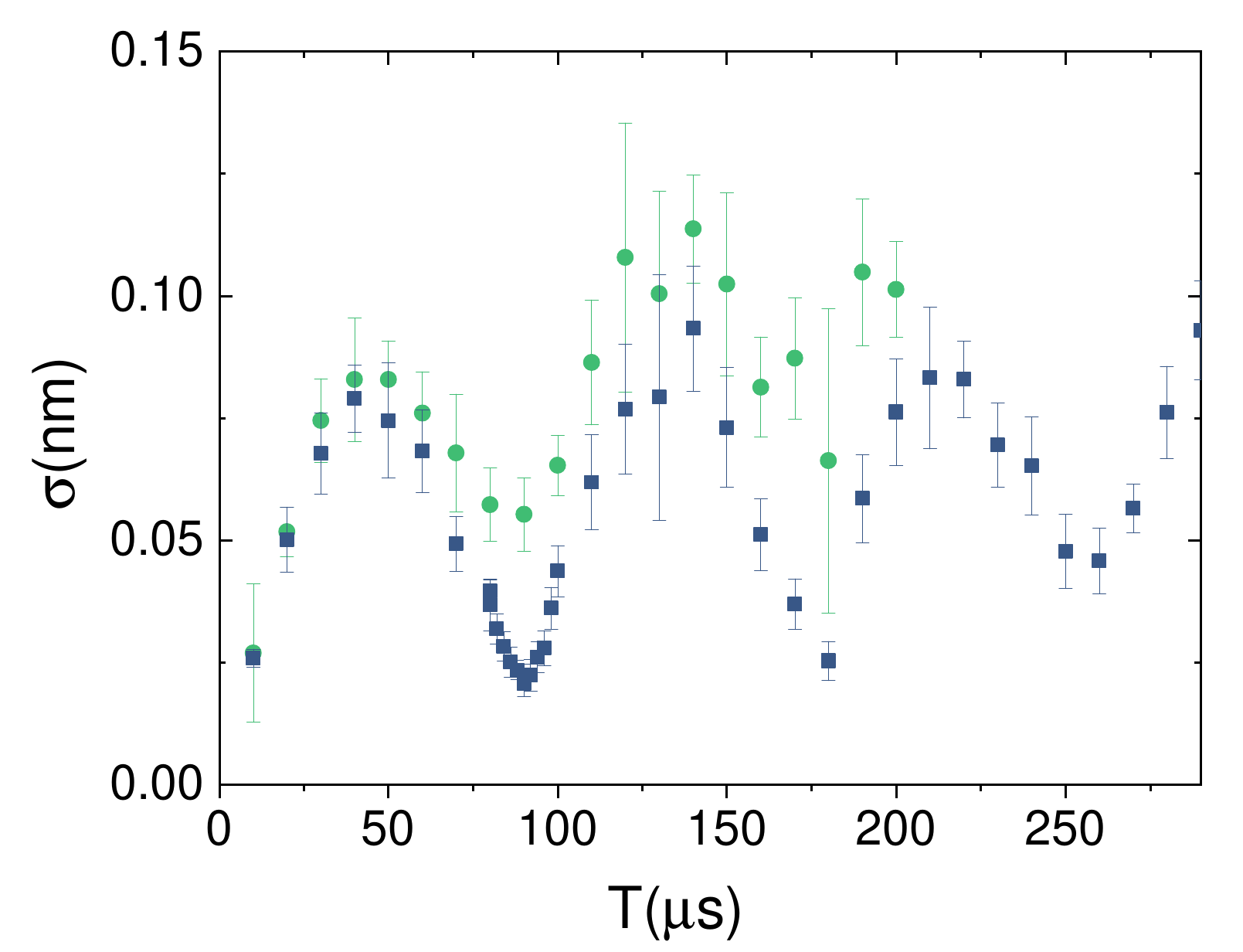}
\caption{
\textbf{Influence of vibrational noise on heating.}
Time evolution of the position variance $\sigma$ for two pumping configurations. The square and circle points show the condition with an ion pump and a titanium sublimation pump and that with a turbo-molecular pump and  a scroll pump, respectively. The scroll pump adds significant noise on the experimental setup and faster heating when the laser intensity is lowered.
}
\label{fig:vibration}
\end{figure}

\bibliographystyle{apsrev}

%\bibliography{references,NPbib,ultracold}

\end{document}